\documentclass[%
 aip,
 amsmath,amssymb,
 reprint,%
]{revtex4-1}

\usepackage{graphicx}
\usepackage{dcolumn}
\usepackage{bm}
\usepackage{tabularx} 

\usepackage[utf8]{inputenc}
\usepackage[T1]{fontenc}
\usepackage{mathptmx}

\begin{document}

\preprint{AIP/123-QED}

\title{Multiple hit reconstruction in large area multiplexed detectors}

\author{B. Radics}
 \email{bradics@phys.ethz.ch}
\author{G. Janka}%
\affiliation{
Institute for Particle Physics and Astrophysics, ETH Z\"urich, CH-8093 Z\"urich, Switzerland
}%

\author{D. A. Cooke}
\affiliation{%
Department of Physics and Astronomy,
University College London,
Gower Street,
London
WC1E 6BT
UK
}%

\author{S. Procureur}
\affiliation{%
Irfu, CEA, Universit\'{e} Paris-Saclay, 91191 Gif sur Yvette, France
}%

\author{P. Crivelli}%
\affiliation{
Institute for Particle Physics and Astrophysics, ETH Z\"urich, CH-8093 Z\"urich, Switzerland
}%

\date{\today}

\begin{abstract}
A novel approach is presented to unfold particle hit positions in tracking detectors with multiplexed readout representing an under determined system of linear equations. The method does not use any prior information about the hit positions, the only assumption in the procedure is that isolated hit signals generated on consecutive detector strips follow a smooth distribution. Ambiguities introduced by charge sharing from multiplexing are reduced by using a regularization technique. We have tested this method on a multiplexed 50x50 cm$\textsuperscript{2}$ Micromegas detector with 1037 strips and only 61 readout channels, using cosmic rays, and we have found that single and multiple clusters of hits can be reconstructed with high efficiency. 
\end{abstract}

\maketitle

\section{\label{sec:Intro}Introduction}

Since the invention of the wire chamber \cite{Charpak1968} and the subsequent development of Micro-Pattern Gaseous Detectors (MPGDs), modern particle physics experiments routinely employ gaseous tracking sensors \cite{Abbon:2007pq,Iakovidis:2013bsa,Banerjee:2017mdu,ASACUSA2014,GBAR2015} thanks to their ease of use and robustness combined with excellent spatial resolution \cite{Bressan:1998uu,Derre:2000vf}. However, scaling such systems to a large area requires a significant increase in the number of electronic readout channels, which demands high costs. The recent innovative technique of genetic multiplexing \cite{PROCUREUR2013}, developed for Micromegas \cite{Giomataris199629}, offers the possibility of reducing the number of readout channels in large area detectors. This multiplexing approach groups detector strips together in such a way that the redundancy in the signal matches the loss of information due to the charge sharing exactly. An apparent limitation of this readout scheme is the rise in the level of ambiguities when multiple particles hit the same detector plane and the produced charges are shared among some of the detector strips. In this case, the level of multiplexing needs to be fine-tuned in order to reduce the probability of ambiguities to an acceptable level (depending on the incoming particle flux and the physics needs). 

In this work we investigate the possibility of directly unfolding hit positions of multiple particles in MPGD detectors with genetic multiplexing readout, addressing the potential multi-hit ambiguities. As a result, a numerical minimization approach is proposed that seeks for a solution without any prior information on the true hit positions but with a constraint reducing the existing ambiguities. First, we briefly describe the genetic multiplexing readout scheme. Then we introduce the unfolding technique and illustrate it by applying it to simulated data. Finally, we assess the performance of the method using cosmic ray data collected with multiplexed, 50x50 cm$\textsuperscript{2}$  Micromegas detectors  \cite{GIOMATARIS2006405}.
\section{\label{sec:GeneticMultiplex}Genetic multiplexing readout}

The genetic multiplexing scheme was invented by Procureur et al.\cite{PROCUREUR2013}. In this readout solution one arranges the readout channels and strips in the following way: having a multiplexing factor, $m$, and a number of readout channels, $p$ (a prime number), one generates $m$ sublists for the ordering of $m\times p$ strips. For each group of $p$ strips, the ordering is given by the following formula,
\begin{equation}
\mathcal{O} = 1 + [(i
\times s) \mathrm{mod}\: p]    
\label{eq:GenMult}
\end{equation}
where $i$ ranges from $0$ to $p-1$, and $s$ is the $s$th out of the total $m$ sublists. The consecutively printed strips are then grouped together in $m$ groups, and within each of the groups, the strips' connections to readout channels are following the ordering rule given by the generated lists in Eq.\eqref{eq:GenMult}. With such a readout scheme the loss of information from the grouping of strips coincides with the redundancy in the signal given by the special ordering rule. As mentioned before, the presence of multiple hits increases the level of ambiguity for the reconstruction of true hit positions. In the following section, we discuss a possible approach of adding a constraint to the allowed configurations of strips that could produce a particular readout pattern. The particular choice of the constraint removes some of the ambiguities that arose from the grouping of the channels.

\section{Unfolding hit positions with regularization}
In the context of the current work, unfolding refers to the general idea of estimating unknown vector components, $\vec{x} \in V$, of an underlying linear vector space, $V = \mathbb{R} \textsuperscript{n}$, using vectors in the measurement space, $\vec{b}  \in W$, $W =  \mathbb{R}\textsuperscript{p}$, and knowing the (forward) mapping between the two vector spaces, $f: \mathbb{R} \textsuperscript{n} \rightarrow \mathbb{R}\textsuperscript{p}$. For an unmultiplexed detector, $n = p$, while for a multiplexed detector $n > p$. The mapping, $f$, is given by the genetic multiplexing algorithm, and can be represented by a matrix, $A_{p \times n}$, with constant matrix elements. Then one can formulate the unfolding problem by the following classical system of linear equations,
\begin{equation}
A\vec{x} = \vec{b} ,
\label{eq:BasicEq}
\end{equation}
where the coefficient matrix $A$ is known, $\vec{b}$ is measured, and one seeks for a solution for the unknown $\vec{x}$. Since $n > p$, there are fewer equations than unknowns, which leads to an under determined system with infinitely many solutions. For the simplest cases of the genetic multiplexing problem, we observe that the rank of the coefficient matrix and that of the augmented matrix both equals to $p$, which means that the system must have at least one solution. This allows the possibility of introducing a constraint to select the most interesting one out of the possibly infinitely many solutions. A popular choice is Tikhonov regularization \cite{Tikhonov,Calvetti}, where the problem is turned into a minimization problem combined with a regularization term in order to give preference to solutions with particular properties. One then minimizes the following quantity,
\begin{equation}
\min_{\vec{x} \in \mathbb{R} \textsuperscript{n}} \{ \| A\vec{x} - \vec{b}\| \textsuperscript{2} + \lambda^{2} \| L\vec{x}\| \textsuperscript{2} \}
\label{eq:Chi2Min}
\end{equation}
where $L\vec{x}$ is a term, which penalizes unwanted solutions such that $L$ is a matrix acting in the solution space, and $\lambda$ is a tuning parameter controlling the amount of penalization applied during minimization. There is some freedom on how to choose $L$. In the following, we show that the choice of seeking a smooth solution guides us to use the second order difference operator. Then we make use of a direct numerical minimization to find the best estimate for $\vec{x}$.

\section{Performance with simulated data}
\subsection{Single hit reconstruction}

\begin{figure}[htp]
\centering
\includegraphics[height=0.25\textwidth]{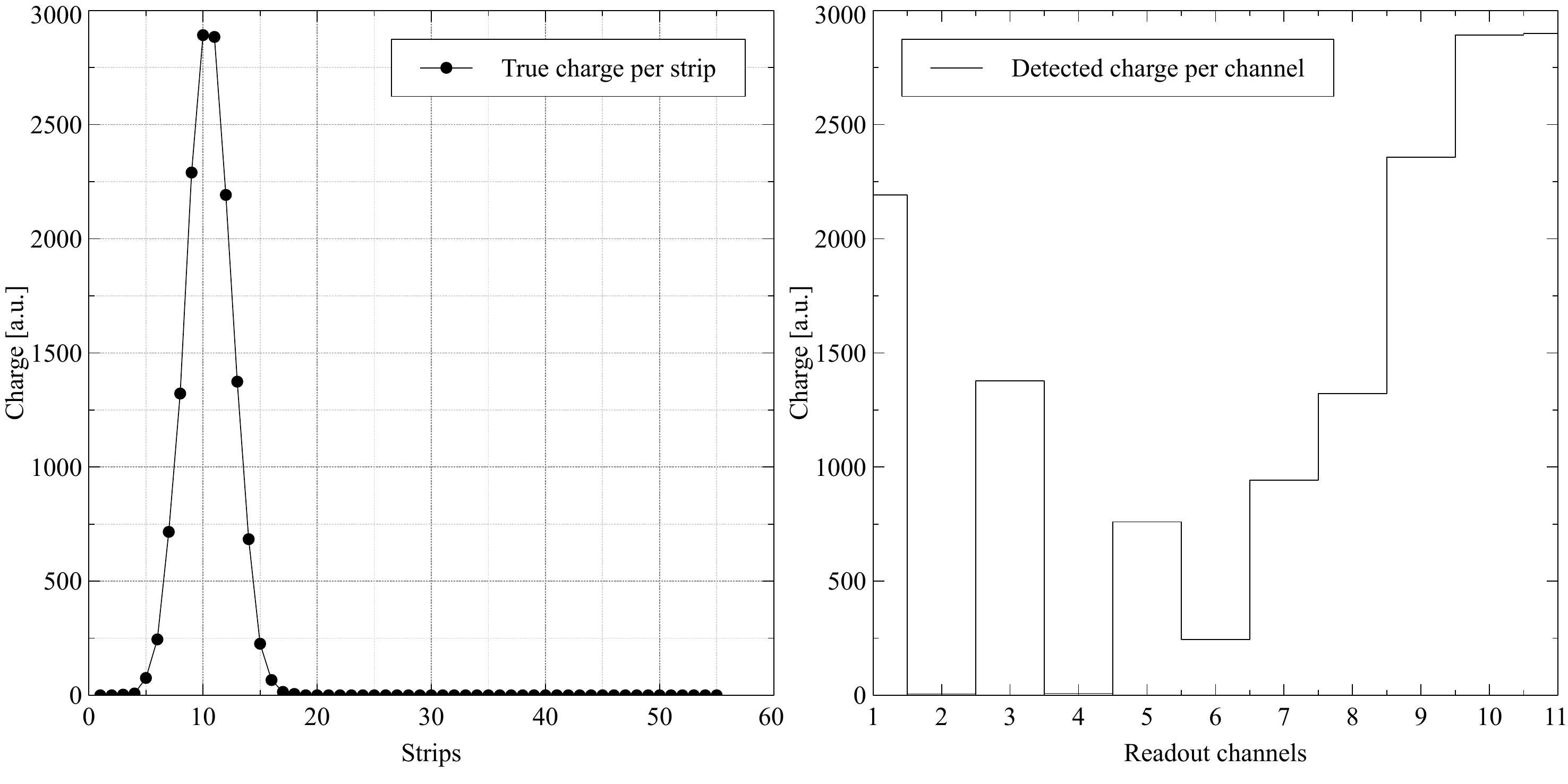}
\caption{Generated toy data: true charges per strip (left) and charges per readout channel after genetic multiplexing (right). }
\label{fig:Toy_data}
\end{figure}

\begin{figure}
\centering
\includegraphics[height=0.4\textwidth]{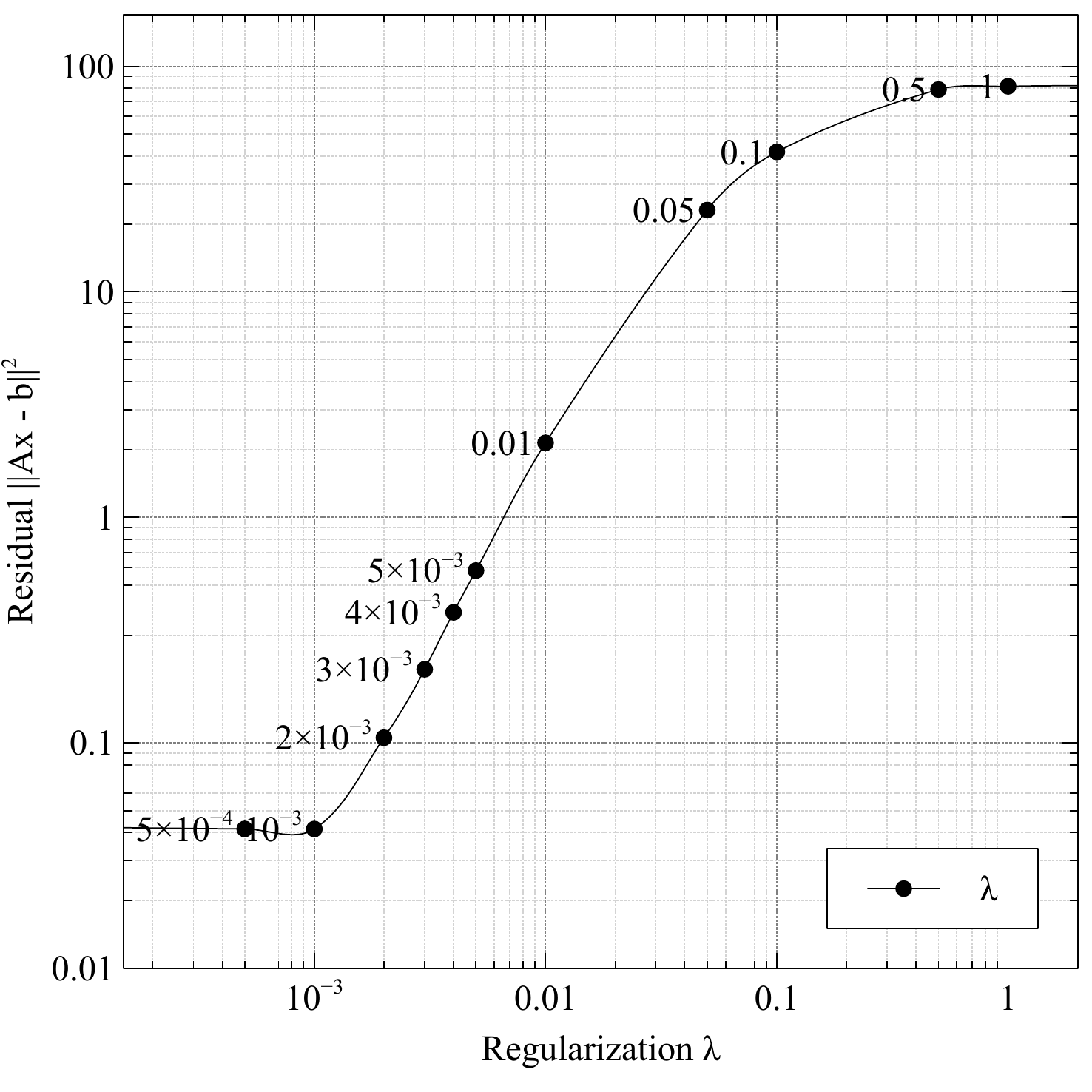}
\caption{Residual charges, $||A\vec{x} - \vec{b}||^{2}$, for various values of the regularizaton parameter, $\lambda$, for the unfolding performed on a system with multiplexing factor, $m = 5$, and number of readout channels, $p = 11$. }
\label{fig:Toy_lambda_scan}
\end{figure}

\begin{figure}
\centering
\includegraphics[height=0.4\textwidth]{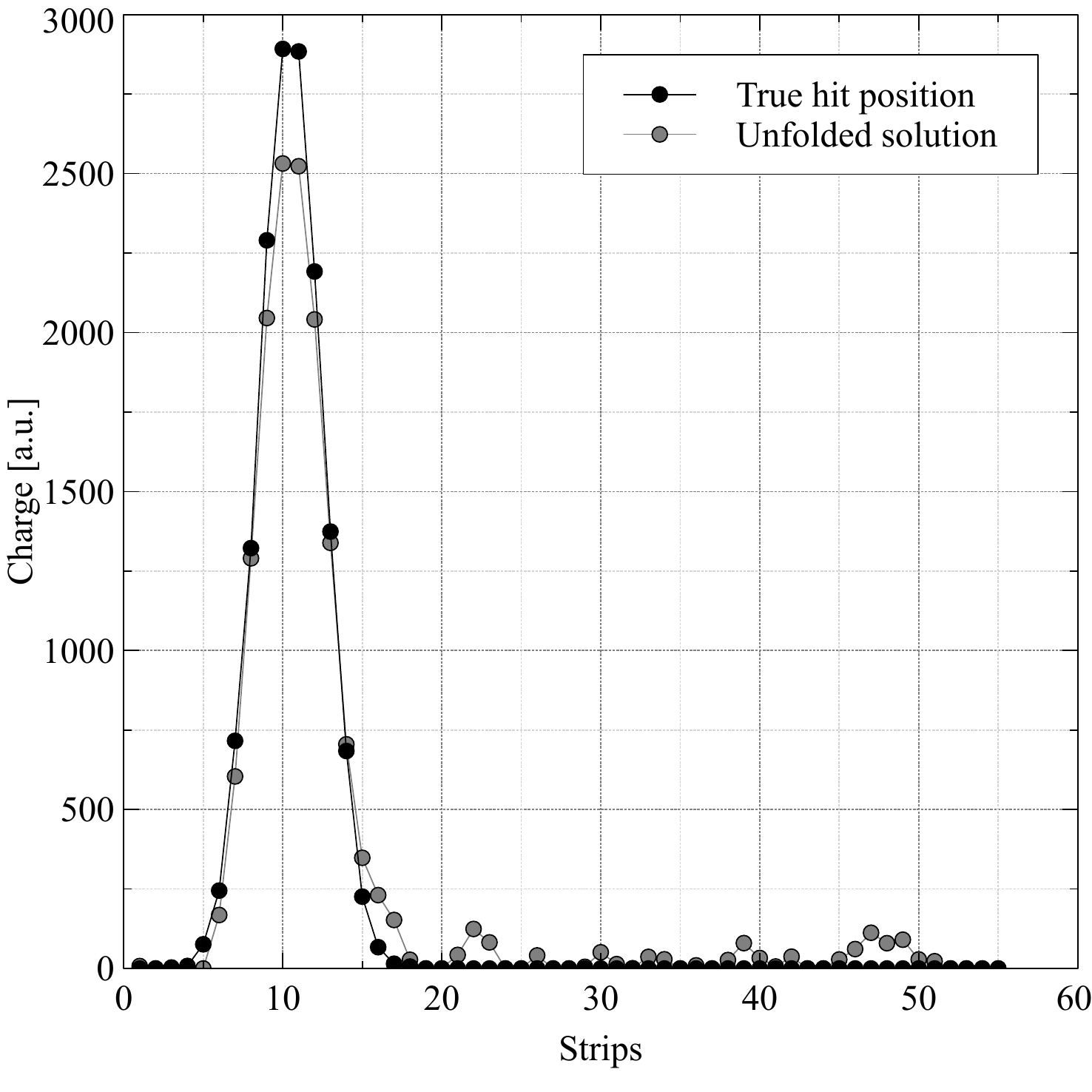}
\caption{True (black) and unfolded (gray) distribution of strip charges for a simulated event with a single cluster, using $p = 11$ readout channels and multiplexing factor $m=5$. }
\label{fig:Toy_Ex1_unfold}
\end{figure}

\begin{figure}
\centering
\includegraphics[height=0.4\textwidth]{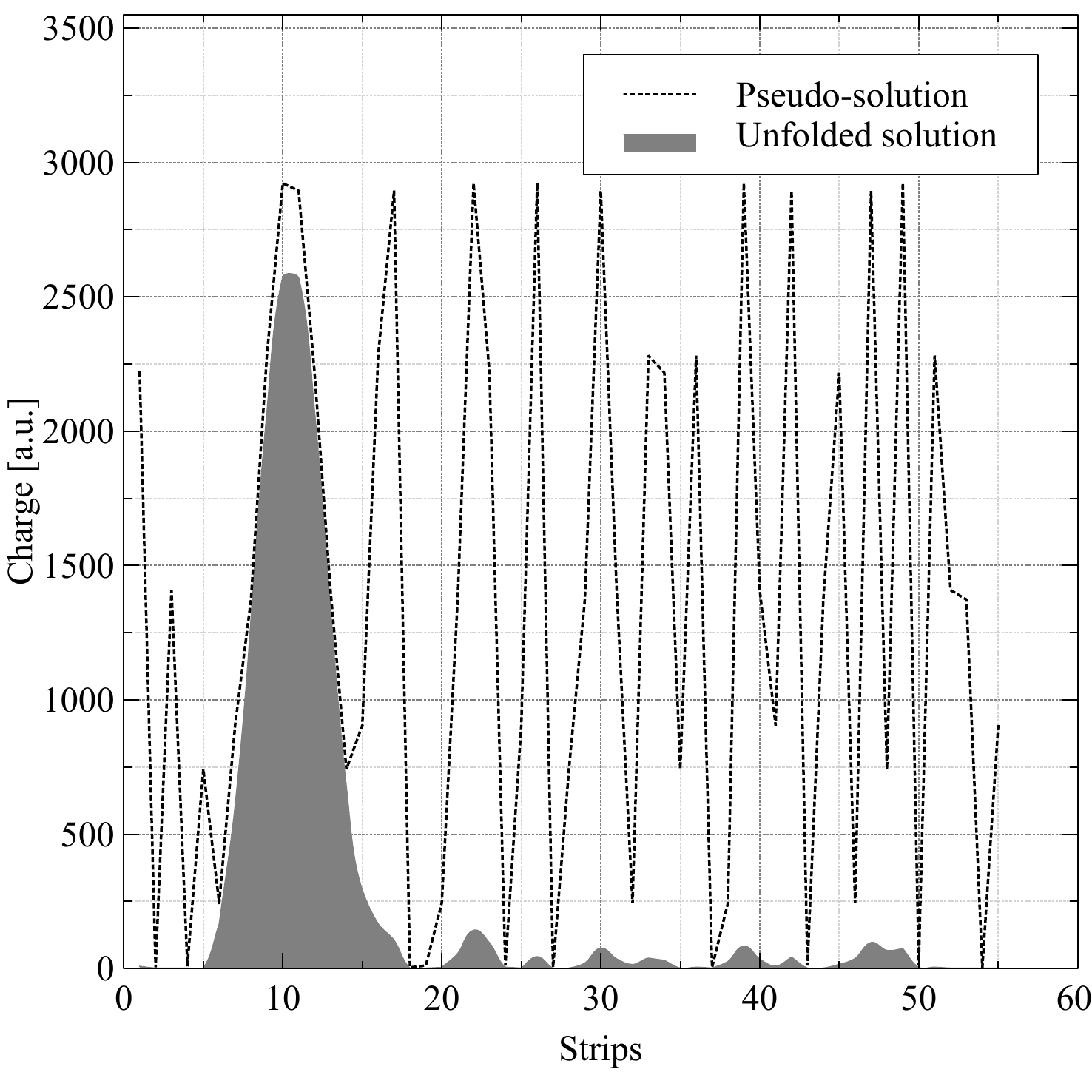}
\caption{The pseudo-solution $A^{T}\vec{b}$ (black, dotted line) and the unfolded (gray, shaded area) distribution of strip charges for a simulated event with a single cluster, using $p = 11$ readout channels and multiplexing factor $m=5$. }
\label{fig:Toy_Ex1_unfold_wPseudoSol}
\end{figure}

In order to investigate the performance of our approach and for illustration purposes, we start with a low dimensional problem, and later we increase the size of our simulated detector. Idealized clusters of signals are generated with a Gaussian shape at various mean strip positions, $\vec{x} = Gaus(\mu, \sigma)$, on $n$ fictitious detector strips, and using $p < n$ readout channels. We generate the multiplexing matrix, $A_{p \times n}$, using the genetic multiplexing algorithm, Eq. \eqref{eq:GenMult}, with number of readout channels $p = 11$, multiplexing factor $m = 5$, and consequently number of strips $n = p \times m = 55$. Then the multiplexing matrix is applied on the generated signal to obtain a simulated measurement, $\vec{b} = A\vec{x}$. An example of simulated true signal distribution, $\vec{x}$, along the $n$ strips, and that of the corresponding measurement readout, $\vec{b}$, in the $p$ channels are shown in Fig.\ref{fig:Toy_data}. 
As a next step, we use the Minuit library \cite{Minuit2} to numerically minimize Eq. \eqref{eq:Chi2Min} and try to recover an estimate for the true $\vec{x}$ at the obtained minimum. The free parameters of the minimization are the charges on the strips, $\vec{x}$.  During the minimization we have tried using various well-known, widely-used standard regularization matrices: the identity matrix, first and second order finite difference matrices, in order to penalize solutions with a large amplitude, or large first or second derivatives, respectively. The best results were obtained when using the second-order finite difference matrix for regularization, 
\begin{equation}
L = 
\begin{pmatrix}
2 & -1 & 0 & \cdots & 0 \\
-1 & 2 & -1 & \cdots & 0 \\
0 & -1 & 2 & \cdots & 0 \\
\vdots  & \vdots &  & \ddots & \vdots \\
0 & 0 & \cdots & -1 & 2 \\
\end{pmatrix}
\end{equation}
which has no effect on smooth underlying signals, while damping solutions with many oscillations. Therefore the regularization improves the reconstruction if the underlying signal has smooth, distinct features. A good detector should produce a contiguous cluster of signals on the consecutive strips, as a result of a passage of a particle through a sensitive volume, therefore the choice of the second-order finite difference matrix is reasonable. At the same time, initially, there is no prior choice on the $\lambda$ parameter, which controls the relative amount of regularization. A standard way of tuning $\lambda$ is to scan with its value and study the residual between the found solution and the input vector, $r_{\lambda} = \|A\vec{x}_{\lambda} - \vec{b}\|^{2}$. A result of such a scan for the current simulated data is shown in Fig.\ref{fig:Toy_lambda_scan}. Large $\lambda$ values were found to lead to solutions with preference to over-smoothing, i.e. smoothing out even important characteristics of the underlying signal. Very small $\lambda$ values indicate a solution without any penalization on smoothing, giving solutions with large oscillations. A good compromise was identified using $\lambda$ values which produce small residuals but still with sufficient penalization over oscillations. Such a  choice is $\lambda \simeq 2-5\times 10^{-3}$, which lies close to the corner of the curve at $\lambda \simeq 10^{-3}$, below which the smoothing has no effect any longer. It is noted that the particular choice of $\lambda$ is only valid for the genetic multiplexing parameters used in this example, $p=11$ and $m=5$.
An typical result of unfolded hit positions is presented in Fig.~\ref{fig:Toy_Ex1_unfold}, where we show the comparison of the true underlying signal with the one obtained from Minuit. The signal cluster is successfully reconstructed, however, there are small residual charges in the solution. To understand their origin we show on Fig.~\ref{fig:Toy_Ex1_unfold_wPseudoSol} the same solution together with the \textit{pseudo-solution}, which is the vector of charges calculated as $A^{T}A\vec{x} = A^{T}\vec{b}$, where $A^{T}$ denotes the transpose of the multiplexing matrix. This transformation projects into the vector space of $\vec{x}$. The pseudo-solution contains the true cluster solution as well as a number of fake clusters. The latter are suppressed by the regularization. However, the small residuals align with the fakes in the pseudo-solution and therefore they are consistent with the suppressed fake clusters.

\begin{figure}
\centering
\includegraphics[height=0.4\textwidth]{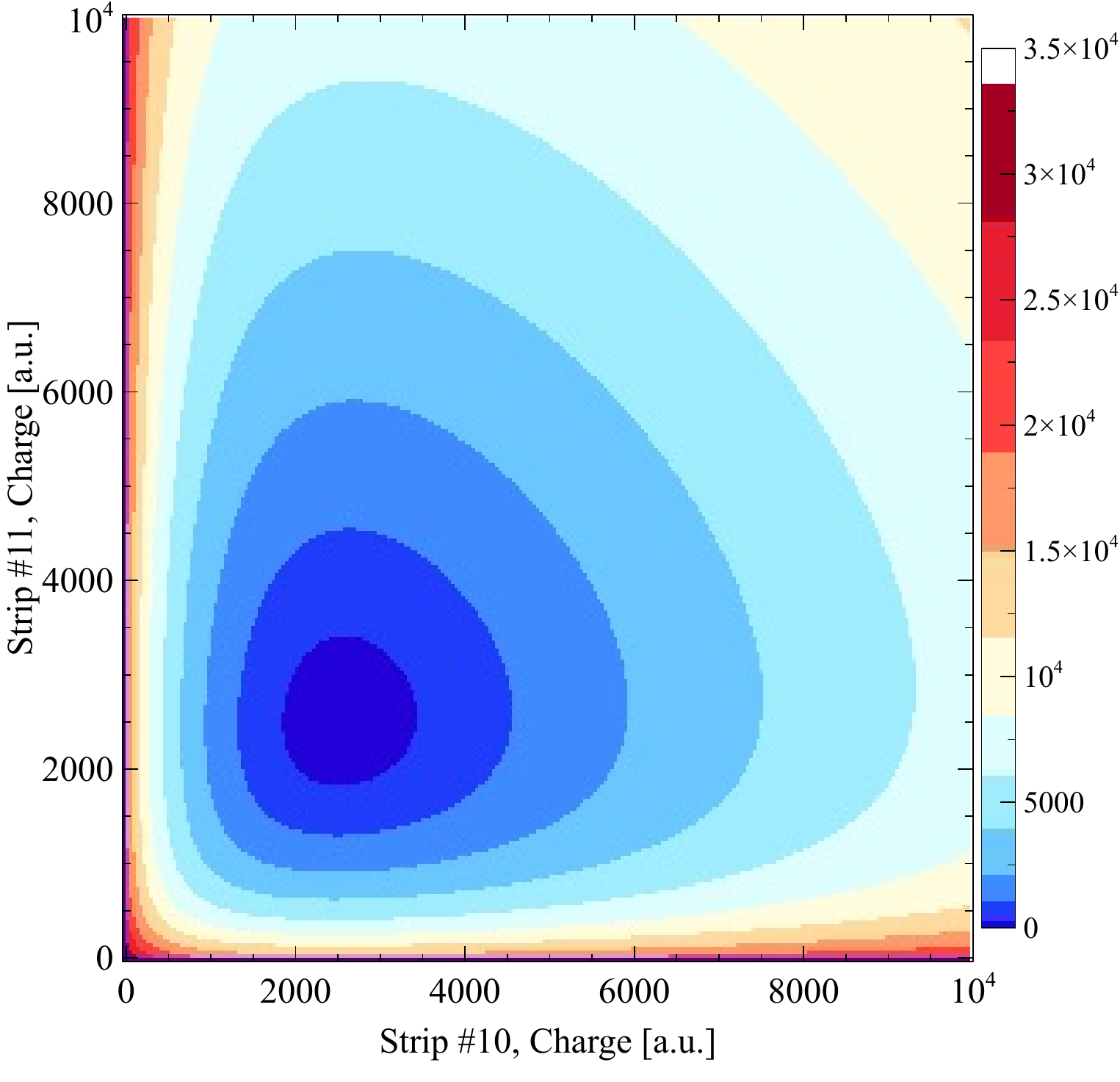}
\caption{Two-dimensional scan of strip charge values around the maximum of the hit position (strip \#10 and \#11) for the solution shown in Fig. \ref{fig:Toy_Ex1_unfold}. The color scale indicates the value of the target minimization function. }
\label{fig:Likelihood_scan}
\end{figure}

In order to demonstrate that Minuit finds good minima during the processing, in Fig.\ref{fig:Likelihood_scan} we show the target quantity in Eq. \eqref{eq:Chi2Min} as a function of two strip charges values around the maximum amplitude. The minimum in the regularized target function evidently points to charge amplitudes covering the true values.

\subsection{Multiple hit reconstruction}
\begin{figure}
\centering
\includegraphics[height=0.45\textwidth]{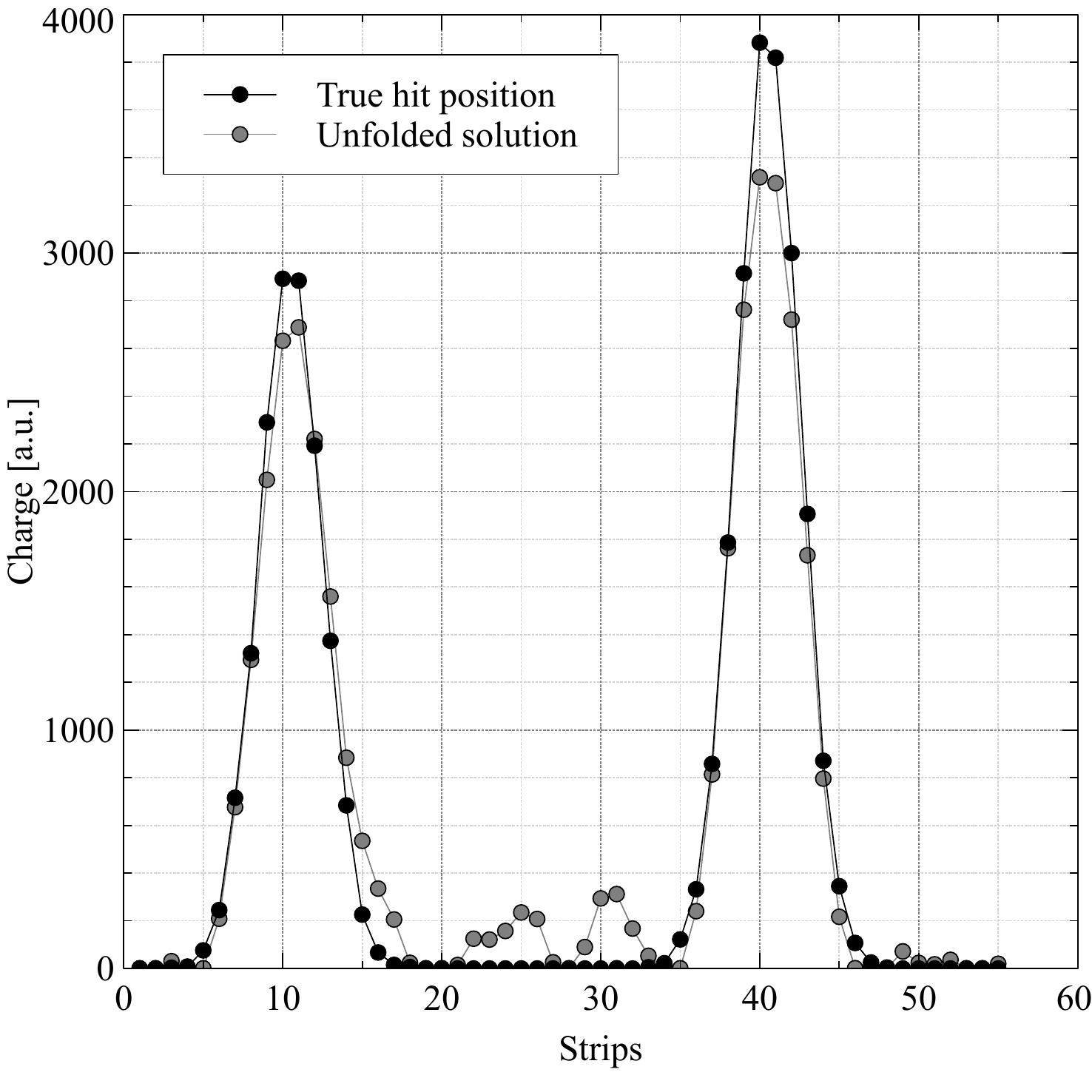}
\caption{True (black) and unfolded (gray) distribution of strip charges for a simulated event with two clusters, using $p = 11$ readout channels and multiplexing factor $m=5$. }
\label{fig:Toy_Ex2_unfold}
\end{figure}

The performance of the unfolding method is illustrated in Fig. \ref{fig:Toy_Ex2_unfold} for the case of simulated double hits. In this case, the small, residual oscillations are somewhat more pronounced because as more channels fire there is also more possibility for fake clusters to exist as solutions. They, however, do not affect the hit position reconstruction, and may be treated by further post-processing of the solution. 

\subsection{Reconstruction with large multiplexing factor}
\begin{figure}
\centering
\includegraphics[height=0.45\textwidth]{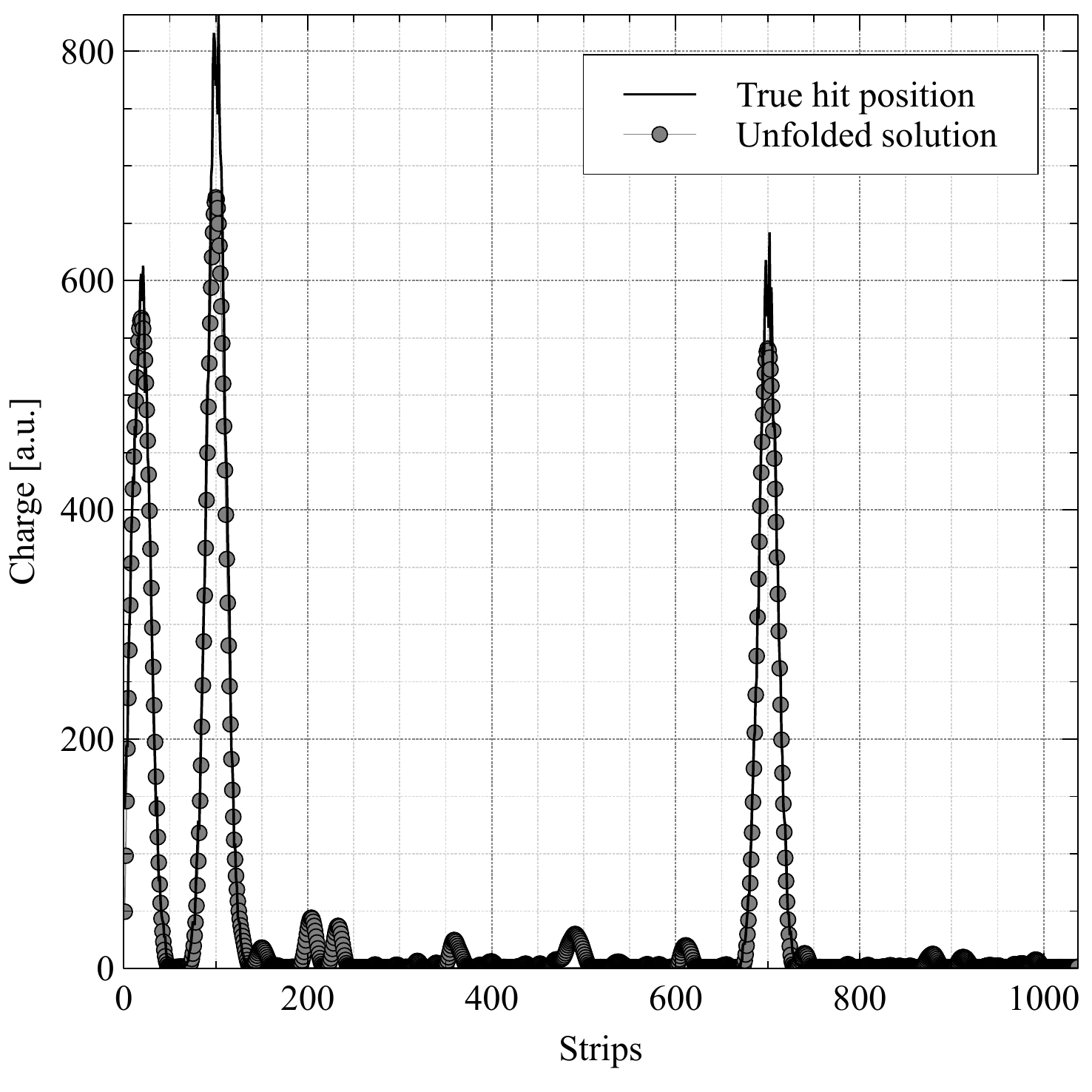}
\caption{True (black) and unfolded (gray) distribution of strip charges for a simulated event with three clusters, using $p = 61$ readout channels and multiplexing factor $m=17$.}
\label{fig:Toy_Ex3_unfold}
\end{figure}

As a final example, we present the performance of the method on a problem with a large number of dimensions: multiplexing factor $m = 17$, and readout channels $p = 61$, with $n = 17\times 61 = 1037$ strips. This means that Minuit has to find minima in the (maximally) 1037-dimensional parameter space. In practice, however, the dimension of the parameter space is lower because the mapping of the strips to readout channels is known, therefore inactive channels can be used to eliminate strips from the solution space in advance. For such a large dimensional problem we had to retune the regularization parameter, $\lambda$, and the best value was found to be $\lambda \simeq 0.5$. This may be interpreted in such a way, as previously discussed, that for larger systems and more channels fired there are more possibilities to form fakes, hence these cases may need a larger amount of penalization to find a smooth solution. The result is shown in Fig. \ref{fig:Toy_Ex3_unfold}. All the positions of the multiple hits have been correctly recovered. As seen before, the unfolded amplitude of the hit signals are slightly lower than the those of the true hits, however, this does not impact the reconstruction of the mean position and general shape of the hit clusters, which is the main interest for tracking detectors.

\subsection{Closed-form solution}
One particular strength of Tikhonov regularization is the existence of a solution in closed form, which could replace the iterative minimization step in principle. Generally, this can be written as:
\begin{align}
\vec{x} = \left(A^TA + \lambda^2L^2\right)^{-1}A^T \vec{b}\label{eqn:cf}
\end{align}
with the variables defined as earlier. For the particular problem of unfolding multiple hit positions in large detectors with high multiplexing factor, however, it is often the case that the resulting matrix to be inverted has high condition number (i.e. high sensitivity of the solution, $\vec{x}$, to small changes in the input, $\vec{b}$), leading to poor solutions. Therefore, we decided to use the Minuit numerical minimization approach for further data analysis, which gives more robust results.
 
\section{Performance with cosmics data}
We have used cosmic data to verify that the unfolding approach works as well for real detectors. In particular, the main aim of this work was to investigate the unfolding performance in the presence of multiple hits per detector layer. Therefore, we first selected single-track events from the cosmic data and used those to artificially produce mixed, multiple hit events. This allows quantifying the efficiency of the hit finding for the case of multiple hits.

\subsection{The Micromegas detector}
The Micromegas detectors used in this work were built at Saclay for muography projects \cite{muography} and as a prototype tracker system for the GBAR experiment \cite{Dipanwita:2017}. They have an active area of 50x50\,cm$^2$, for a total length of 54.6\,cm, a 2D readout with 1037 X (horizontal) strips and 1037 Y (vertical) strips arranged in 2 layers within the Printed Circuit Board (PCB), a Kapton layer with resistive strips between the micro-mesh (bulk technology) and the readout channels. The readout was implemented following the genetic multiplexing, with a multiplexing factor of $m$ = 17, resulting in 61 channels (1 connector) for each coordinate of a detector, and a drift gap ranging from 8 to 15\,mm ensured by an Aluminum frame. The detectors were equipped with an electronics readout system based on the DREAM ASIC \cite{dream} developed for large capacitance detectors. The data was taken with 4 such Micromegas detectors arranged to a telescope, measuring in self-triggering mode. The detectors were placed at z=550, 450, 100 and 0 mm respectively along the telescope axis.

\subsection{Data preprocessing}
As a first step, we used our unfolding approach to reconstruct the hit positions of an event in every detector layer. A cluster finder algorithm was used then to identify isolated hits. The mean cluster positions, total charge, and widths were calculated for all the found clusters. In order to suppress noise-induced clusters, we applied the following discrimination cuts: the clusters taken for further processing were required to have a minimum cluster width of four strips and an amplitude larger than 150 ADC counts. Events were then preselected for cases with only one hit per each detector plane, and in addition, we also required that these hits form a straight line track. Only these events were used for further analysis. 

Out of the preselected events with a single track, we created a dataset of double-hit events by merging each pair of events. Merging was performed by adding the collected charges at the readout level channel by channel. That is having $A$ as the multiplexing matrix and $A\cdot \vec{x}_{1}= \vec{b}_1$ and $A\cdot\vec{x}_{2}=\vec{b}_2$ as two measurements, we added directly the measurement vectors, $\vec{b}_\mathrm{merge} = \vec{b}_1 + \vec{b}_2$, which implies that the solution for the merged event should be just the vectorial sum of the true underlying solutions $\vec{b}_\mathrm{merge} = A\cdot(\vec{x}_{1}+\vec{x}_{2})$. All possible combinations of single track events were thus merged together in order to maximize the possible double hit cases. With this dataset, we used again the unfolding approach, along with the same cluster finding and discrimination cuts as mentioned before, in order to find how many times double-hit events were successfully unfolded.

\subsection{Results}
\begin{figure}
\centering
\includegraphics[height=0.45\textwidth]{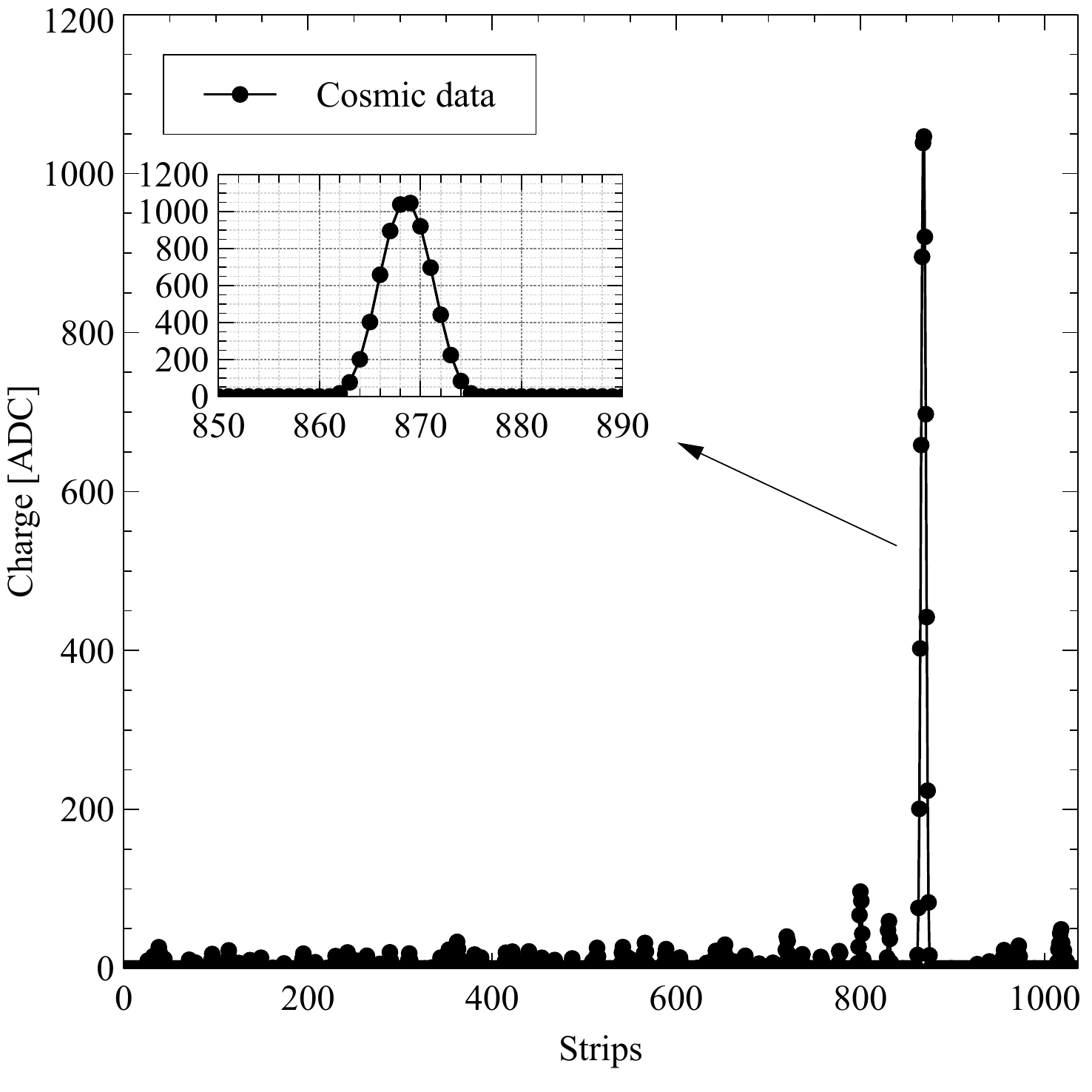}
\includegraphics[height=0.45\textwidth]{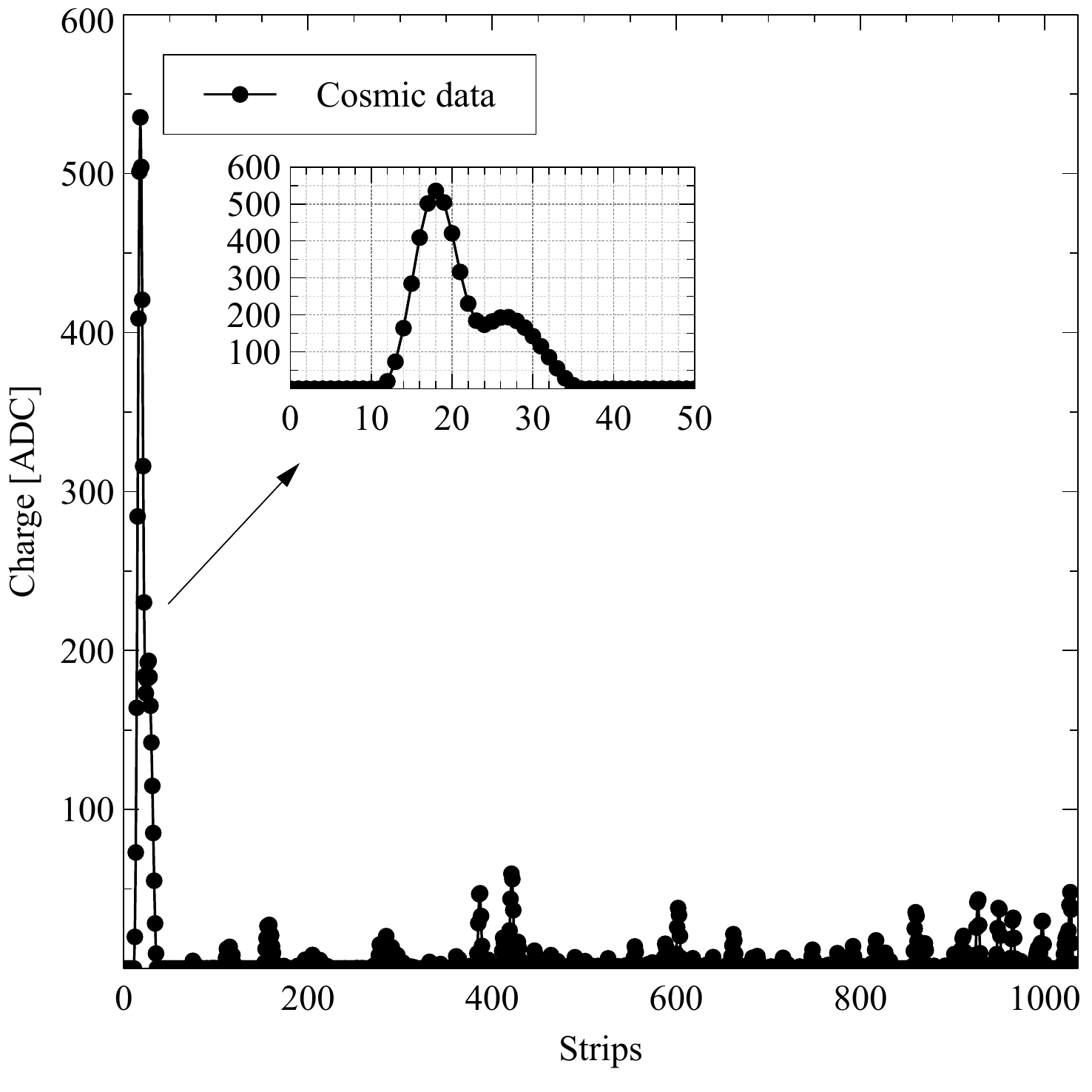}
\caption{Example result of the unfolding method for two cosmic data events (top and bottom), each with a single hit cluster, for 1037 strips with $p = 61$ readout channels.}
\label{fig:Data_ex1_unfold}
\end{figure}

\begin{figure}
\centering
\includegraphics[height=0.45\textwidth]{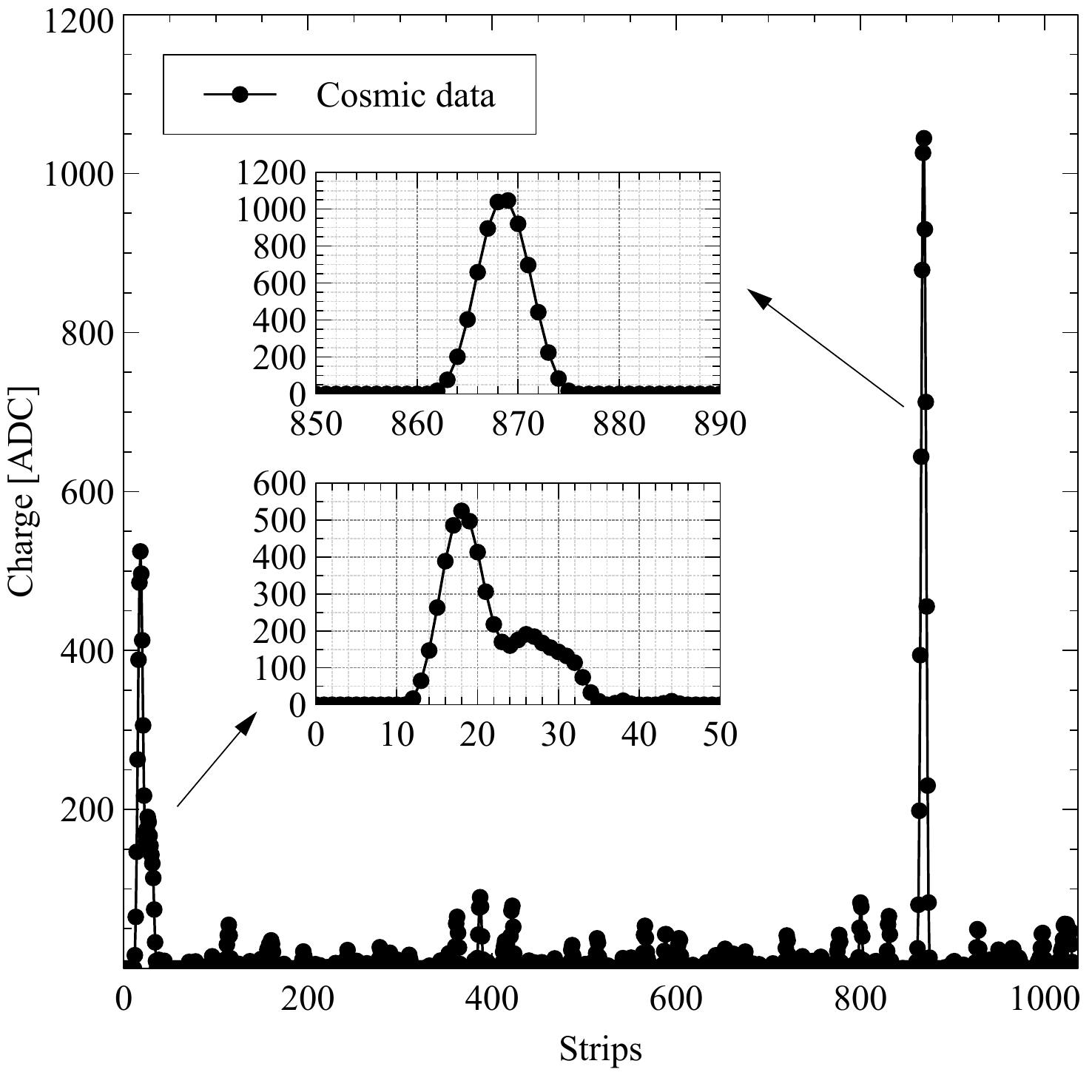}
\caption{Example result of the unfolding on a merged double-hit cosmic event, created from the events shown in Fig.\ref{fig:Data_ex1_unfold}.}
\label{fig:Data_ex2_unfold}
\end{figure}

\begin{figure}
\centering
\includegraphics[height=0.45\textwidth]{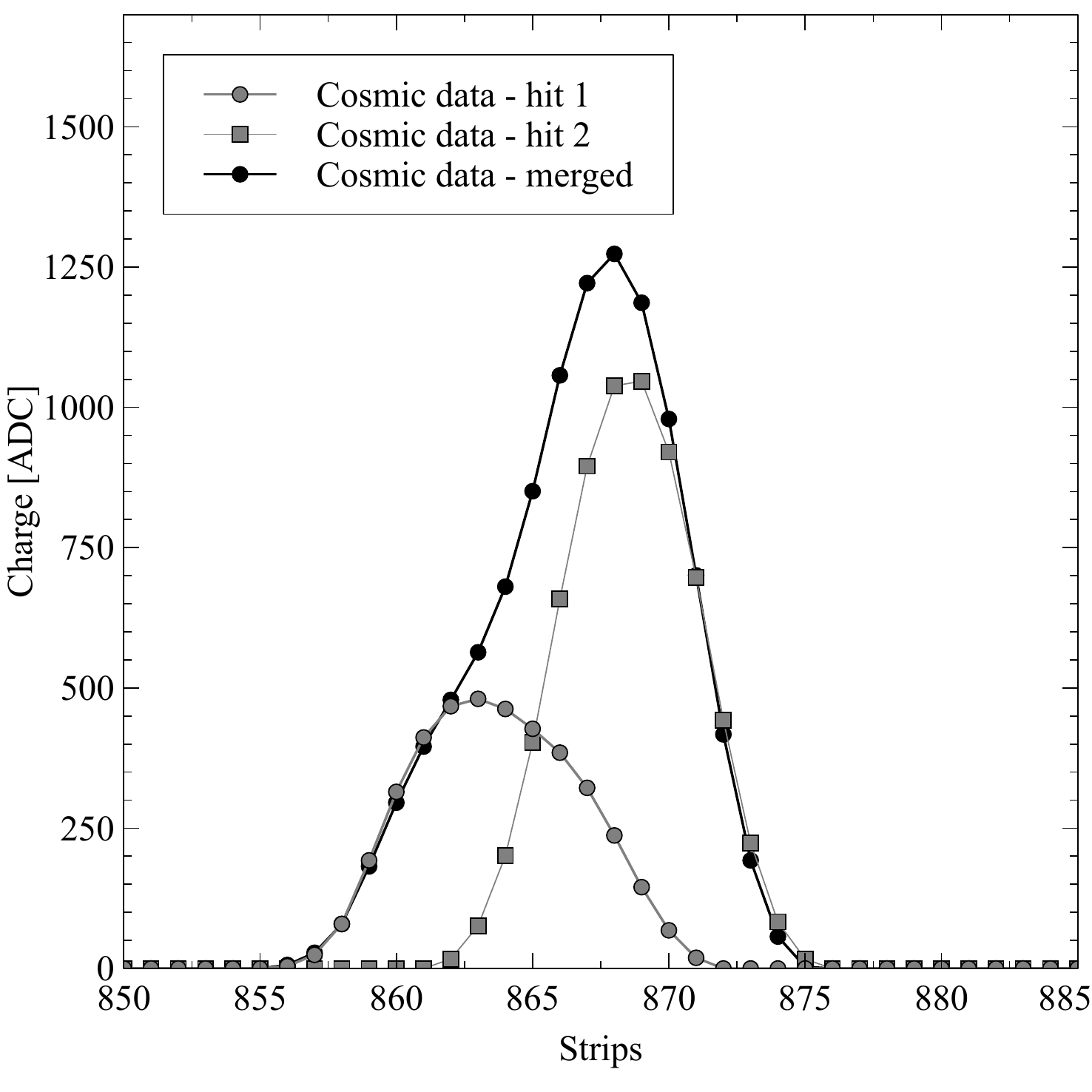}
\caption{Example result of the unfolding when the cosmic events had nearby hits  (grey circle and square). The unfolding method found a smooth solution (black circle) containing the two original single hit cluster shapes.}
\label{fig:Data_closesignal}
\end{figure}

\begin{figure}
\centering
\includegraphics[height=0.45\textwidth]{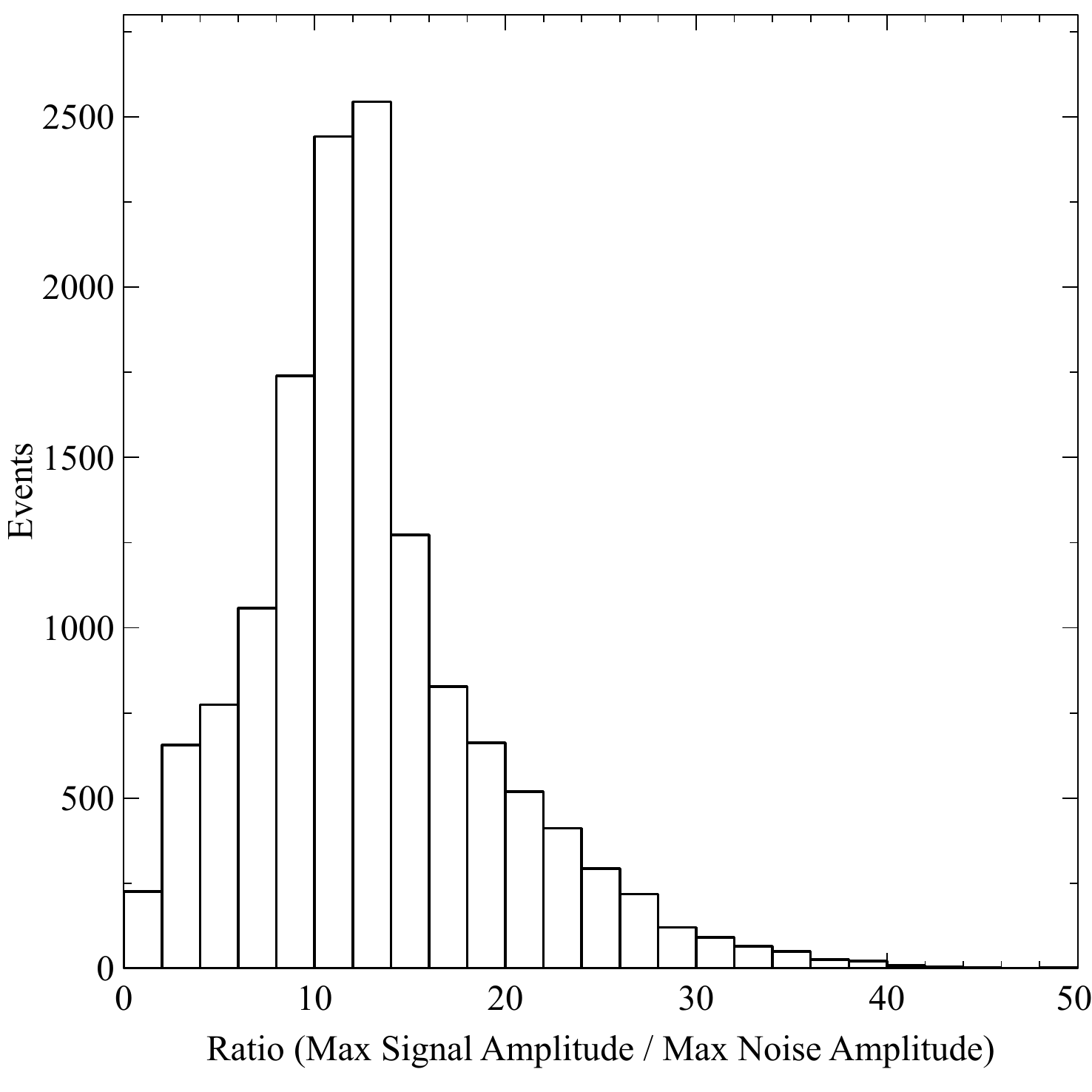}
\caption{Distribution of the maximum signal to maximum residual oscillation amplitude, $S/R$, from single hits in cosmic data.}
\label{fig:Data_wigglesratio}
\end{figure}

The performance of the unfolding method on a Micromegas readout projection for two separate events with single hits in the cosmic data is illustrated in an example in Fig.\ref{fig:Data_ex1_unfold}. The result of the unfolding from the merging of these two events is shown in Fig.\ref{fig:Data_ex2_unfold}. This example demonstrates that our method is capable of resolving the signals of cosmic double hits. Both the pulse heights and the pulse shapes were fully recovered. We note that in the case of Fig.\ref{fig:Data_ex2_unfold} the hit at the lower strip coordinates has a double bump structur. We show a similar example on Fig.\ref{fig:Data_closesignal}, where the cosmic single hit events (grey circles and squares) were selected to be at nearby positions. The unfolding output (black circles) from the merged event covers both of the nearby hits from the single hit events. This shows that the regularization favoring smooth solutions correctly recovers the underlying cluster shapes.
In order to quantify the performance, we applied the unfolding algorithm to the entire dataset and counted when the two original hits have been reconstructed correctly. The results are shown in Table \ref{tab:results_unfold}. We found a double-hit reconstruction efficiency, $\epsilon$, above 90\%. This depends on the ratio of the maximum amplitude of the signal clusters (S) and that of the residual oscillations (R), since the latter could also be misidentified as signal clusters. The distribution of the $S/R$ ratio from single hits in cosmic events is shown in Fig. \ref{fig:Data_wigglesratio}.  

Most of the events were found to have a ratio of $S/R > 10$. When we consider the double-hit reconstruction efficiency as a function of this ratio, already at $S/R > 4$ we get an efficiency of $\epsilon > 94\%$. Most of the inefficiency was found to be originated from events when only one of the hits was found due to the discrimination cut on the minimum cluster amplitude, or from events when the two hits were too close to each other to be resolved (example shown in Fig.\ref{fig:Data_closesignal}).

\begin{table}
\caption{\label{tab:results_unfold}Double-hit reconstruction efficiency as a function of the maximum signal to maximum residual oscillation amplitude, $S/R$, from cosmic data. }
\begin{ruledtabular}
\begin{tabularx}{\textwidth}{dd}
\mbox{\textbf{S/R ratio}} & \mbox{\textbf{Efficiency [$\%$]}}\\
\hline
\mbox{No cut} & \mbox{91.3} \\
\mbox{S/R $\geq 2$} & \mbox{92.8} \\
\mbox{S/R $\geq 3$} & \mbox{94.2} \\
\mbox{S/R $\geq 4$} & \mbox{94.1} \\
\end{tabularx}
\end{ruledtabular}
\end{table}

\begin{figure}
\centering
\includegraphics[height=0.45\textwidth]{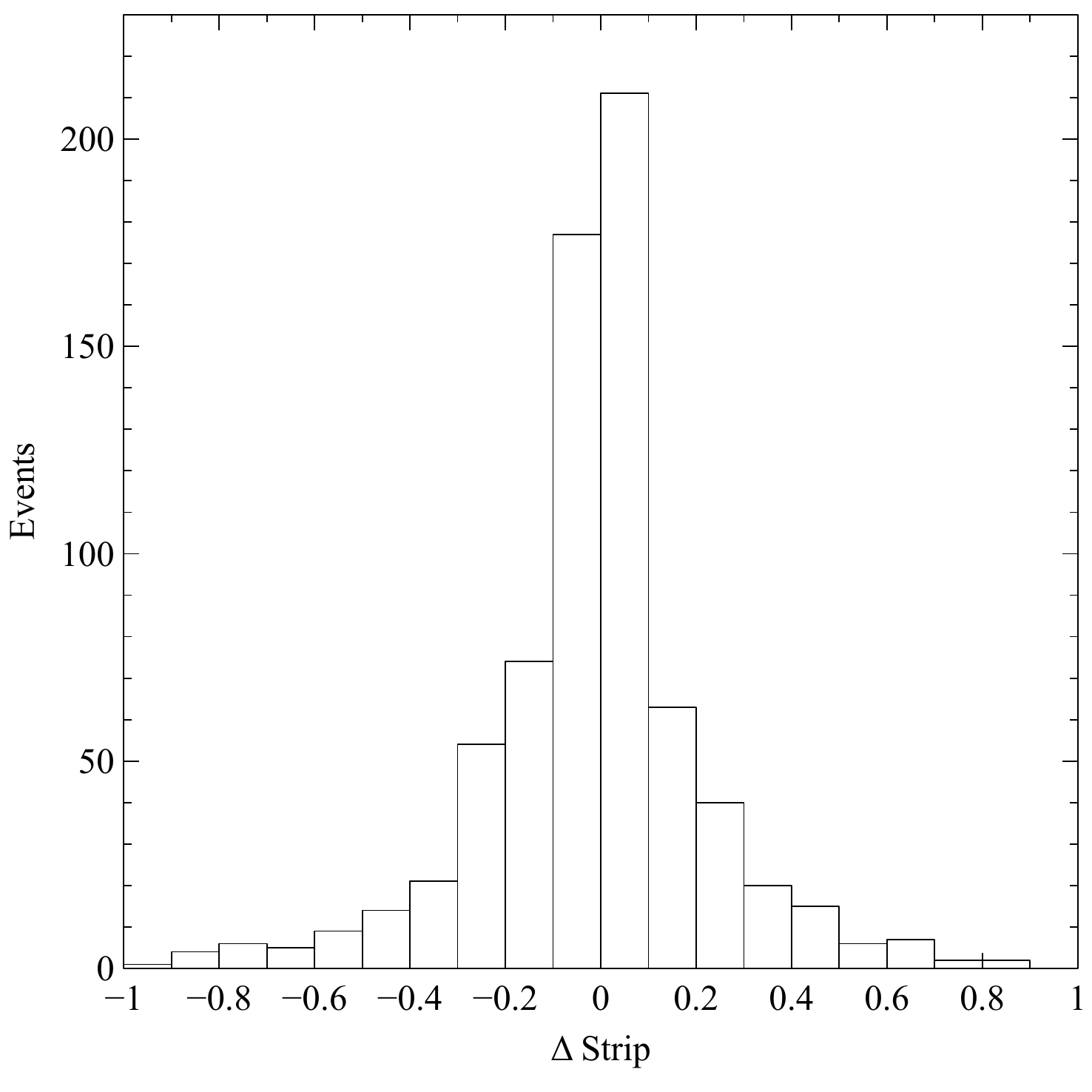}
\caption{Distribution of the difference between the true and reconstructed in mean hit positions from cosmic data.}
\label{fig:Data_posresidual}
\end{figure}

We show the distribution of the difference between the reconstructed and true hit position of the cosmic data events in Fig.\ref{fig:Data_posresidual}, where the difference is given in units of strip coordinates. The true hit position is calculated from the charge-weighted mean from single hit events prior to merging. For all of the events when the hits were found the weighted mean hit coordinates were reconstructed within $\Delta \mathrm{Strip} = \pm 1$ strip.

In the discussion so far the possibility of missing strips (holes) in the signal was not considered. In practice, however, it can occur that there are temporary dead strips in the solution space, which would violate the requirement of a smooth solution, therefore the regularization could break down. We have simulated various cases with a hole and found that the performance of the unfolding algorithm depends on the position of the missing strip. In case the dead strip is in a region of the detector where there is no true signal (or it is in the tail of a cluster) there is no (or hardly any) impact on the performance. The only case where we encountered problems is when the dead strip is close to the maximum amplitude position of the cluster. In this case the solutions show strong oscillatory behaviour, as one would expect, since the true solution should be an oscillating one. This latter case simply adds to the overall inefficiency, falling into the category $S/R \sim 1$.

\section{\label{sec:Conclusion}Conclusions}
In this work, we presented a novel approach to reconstruct multiple-hit events in Micromegas detectors with genetic multiplexing readout. Without any prior assumption on the true hit positions, and with the only requirement of searching for a smooth solution in a high dimensional space, the method was able to correctly recover the true hit cluster shapes and mean position up to a multiplexing factor $p=17$. The performance was evaluated with double-hit events in cosmic data, and a reconstruction efficiency of $\epsilon \geq 91-94\%$ was achieved depending on the signal cluster amplitude. Inefficiency can be accounted for mainly due to unresolved, nearby clusters and residual oscillations from the unfolded solution. With an improved detector gain, the method has a potential for higher reconstruction efficiency. As a result, the genetic multiplexing readout scheme combined with the presented unfolding approach might allow larger area MPGD detectors to be used without a significant increase in the number of readout channels. An important consequence of the current work is that large area multiplexed detectors might be an attractive solution in environments with very high particle flux.

\begin{acknowledgments}
This work was supported by the Swiss National Science Foundation (grant number 20FL21\_173597) and ETH Z\"urich (grant number ETH-46 17-1) .
\end{acknowledgments}

\nocite{*}
\bibliography{Bibliography}

\end{document}